\documentclass{scrartcl}
\usepackage{CJK}
\usepackage{authblk}
\usepackage[left=3cm,right=3cm,top=3cm]{geometry}
\usepackage{amsmath,amsfonts,amssymb,amsthm}
\usepackage[utf8]{inputenc}      
\usepackage{tikz}
\usetikzlibrary{decorations.pathreplacing}
\usetikzlibrary{shapes}
\usepackage{bbm}     
\usepackage{hyperref}
\usepackage{multirow}
\usepackage{algorithmicx,algpseudocode}
\usepackage{graphicx}

\DeclareOldFontCommand{\bf}{\normalfont\bfseries}{\mathbf}

\begin{document} 
	
\title{Multiple Constrained Routing Algorithms
	in Large-Scaled Software Defined Networks\\
}

\author[*]{Chenyang Xu}
\author[*]{Liangde Tao}
\author[+]{Huajingling Wu}
\author[*]{Deshi Ye}
\author[*]{Guochuan Zhang}

\affil[*]{Zhejiang University, China,\authorcr \{xcy1995, tld, yedeshi, zgc\}@zju.edu.cn}
\affil[+]{Huawei Technologies Co., Ltd., Hangzhou, China, \authorcr wuhuajingling@huawei.com}
\maketitle


\begin{abstract}

In this paper, we consider the bandwidth-delay-hop constrained routing problem in large-scaled software defined networks. A number of demands, each of which specifies a source vertex and a sink vertex, are required to route in a given network. We are asked to select a subset of demands, and assign a routing path for each selected demand without violating the hop and delay constraints, while assuring that the bandwidth occupied in each edge is not beyond its capacity. The goal is to maximize the throughput (the total bandwidth) of the selected demands. We develop an efficient heuristic algorithm for the problem, which consists of three main steps, namely, computing feasible paths for each demand, sorting the demands with some priority rules, selecting a path for each demand. The algorithm is tested with networks of actual sizes and topologies, generated by Huawei Technologies Company. The experiments show that the proposed approach outperforms existing algorithms both in throughput and in running time. In the experiments, our algorithm achieves more than 90\%
of the total bandwidth of the given demands within 10 seconds. Moreover, a large part of our algorithm can run in parallel which largely speeds up the process when using multi-core processors.
\end{abstract}

\section{Introduction}
Routing algorithms play a key role in many computer networks. Usually, routing decisions based on local selfish choices are far away from that of global optima ~\cite{{roughgarden2002bad,qiu2006selfish}}. Recently, software defined networking (SDN) proposes a new paradigm, which is designed to allow programmers to define and reconfigure the way of routing within the network. SDN separates the control plane and the forwarding plane. In an SDN environment, a central controller maintains a flow table to deal with all requests. The controller has a global view of the current network states. This characteristic makes SDN more convenient to handle a breakdown than traditional networks. As a breakdown comes in an SDN-enabled network, the controller will find which links cannot be used currently and which demands are affected by this breakdown, and it will re-route these demands as quickly as possible to avoid worse influence caused by the breakdown.

In this paper, we consider an optimization problem of routing in an SDN-enabled network, which computes routes for multiple source-destination demands to meet additional QoS requirements. To this end, we consider constraints for bandwidth, delay and hop. These constraints arise in practical scenarios. For example, in multimedia applications, each demand has a different level of bandwidth requirements, and each edge in the network has a transfer delay and a bandwidth capacity due to the physical limitation of the actual optical fiber between the vertices. Moreover, hop constraints are required to improve the real-time performance. The aim of a routing algorithm is to maximize the throughput, i.e., the total bandwidth of the demands routed successfully.

Among various models, the most relevant one is the maximum multi-commodity flow problem, which arranges flows for demands from the source to the destination while satisfying the capacity constraints over the edges. If only integral flows are allowed, Even et al.~\cite{Even_Itai_1975} proved the decision version (decides whether there exist integer flows for all demands) is NP-complete, even for only two commodities and unit capacities. Note that our problem is reduced to the integral multi commodity flow problem if only the bandwidth constraints are taken into account. The NP-completeness in general follows immediately.

Although there has been rich literature concerning routing algorithms in centralized networks, this is the first attempt considering bandwidth, delay and hop constraints at the same time in a large-scaled network.

\subsection{Related Work}

Routing with constraints has been extensively studied. A great deal of effort has been focused on two problems: the bandwidth constrained routing problem and the bandwidth-delay constrained routing problem. These two problems differ in the QoS requirements of demands.

To reduce the bandwidth occupation of the whole network, the minimum hop algorithm~\cite{ma_steenkis_1997} greedily selects a path with the least number of feasible links. To deal with the case that two or more minimum hop paths are available, the widest shortest path algorithm~\cite{qoS_routingmechanisms_1999} picks the path with a maximum available bandwidth among them. Wang and Crowcroft~\cite{wang_crowcroft_1996} pointed out that simply choosing the minimum hop path may constitute obstacles for the subsequent demands. They then proposed the shortest widest path algorithm which prefers the path with a less hop among those with the maximum available bandwidth.

In real-world applications, such as routing on backbone networks, an assumption is often made: all possible ingress-egress (source-destination) node pairs are known before processing and the total number of such pairs is small. Based on the assumption, Kodialam and Lakshman~\cite{Kodialam_Lakshman_2000} proposed the minimum interference routing algorithm which chooses the minimum interference path for each demand in order to avoid bandwidth congestion on critical links. Here, a link is called critical to an ingress-egress node pair if whenever the capacity of the link is reduced by one unit, the available capacity of the ingress-egress node pair also decreases by one unit. The level of link criticality is proportional to the number of ingress-egress node pairs which it is critical to. And the path criticality is the total criticality of all links on the path. It could be understood that a path with lower level of criticality interferes less ingress-egress node pairs. So the minimum interference routing algorithm chooses the least critical path for the current demand to reduce possible interference with future demands. Since the maximum flow is a subroutine used to calculate the available bandwidth between all ingress-egress node pairs, the running time of the minimum interference routing algorithm becomes unaffordable as the network scale grows.

In order to provide bandwidth guaranteed paths for large scale network, Jozsa et al.~\cite{Jozsa_2001} proposed a global path optimization algorithm which contains three main parts: demand sorting, path allocation and path deallocation. They measure the priority of each demand as its bandwidth requirement divided by its hop. The algorithm processes demands one by one in non-increasing order of priorities. Before considering the arrangement for demands, the algorithm updates the weight for each link, which represents the bandwidth usage. And for each demand, the least weighted path is selected. To reduce the influence of choosing inappropriate paths for settled demands, path deallocation was preformed to release those selected paths which contain bottleneck edges. The algorithm can handle bandwidth guaranteed routing on network with 400 nodes and 160,000 demands within a few minutes. Undoubtedly, the running time on very large scale networks is still unbearable.

For bandwidth-delay constraint routing, several algorithms have been proposed. The simple solution called the minimum delay algorithm~\cite{wang_crowcroft_1996} chooses the path with the smallest delay in the network after removing all the links with insufficient residual bandwidth. Based on the minimum interference rule~\cite{Kodialam_Lakshman_2000} used for the bandwidth constraint routing problem, an alternative algorithm called the maximum delay-weighted capacity routing algorithm~\cite{Yang_Muppala_2001} interprets the concept of interference in a different way when considering delay constraints. It computes shortest paths with respect to delay for all ingress-egress node pairs. The bottleneck links of these paths are defined to be critical, and the criticality level of a link is a weight function associated with the number of ingress-egress node pairs for which it is critical to. By applying the extended Dijkstra shortest path algorithm~\cite{Chen_1999}, the least weight path satisfying delay and bandwidth constraints is chosen for each demand. Tomovic and Radusinovic~\cite{tomovic_radusinovic_2016} thought that only considering the bottleneck edges lying on the least delay paths for all ingress-egress node pairs is not enough. So Yen's algorithm~\cite{Yen_1971} is applied for each ingress-egress node pair to compute the candidate path set containing $k$ loop-less paths in non-decreasing order of delay. And the weight function defined on the links is modified accordingly. Similarly as the experimental results of the minimum interference routing algorithm for bandwidth constrained routing, although the maximum delay-weighted capacity routing algorithm can achieve good throughput and blocking ratio on small-scale networks, the running time becomes unbearable as the size of network grows. Due to the huge time complexity of Yen's algorithm which is applied to the $k$-shortest path problem, the running time of Tomovic and Radusinovic's algorithm~\cite{tomovic_radusinovic_2016} is much larger compared to the maximum delay-weighted capacity routing algorithm.

\subsection{Our Results}
In this paper we present an efficient heuristic algorithm to tackle the more general model of ours. This approach achieves a high throughput in a large-scaled network with nearly $10,000$ nodes and $10,000$ demands. Moreover, our algorithm runs very fast, where the running time for such a large instance is within $10$ seconds under our experimental environment.

Basically, our algorithm runs iteratively. Each iteration consists of three phases. In the first phase, we compute for each demand a set of paths that are individually feasible. Considering the large number of demands and the multiple constraints on the paths, this phase could be very time consuming. We present several efficient ways to reduce the running time. In particularly, path computing can be implemented in parallel. In the next two phases, we consider the demands one by one and arrange a path (from the path set got in the first phase) for each demand if there is one that fits the network. Clearly, the order of demands and the selection rule of paths are crucial. Inspired by the primal-duality we design a few effective rules for sorting demands and selecting paths. The network is updated as a path is arranged. A new iteration starts if there are demands satisfied in the last iteration and there are unsatisfied demands left. In the experiments, our algorithm is terminated in just a few iterations.

The rest of the paper is organized as follows. Section~\ref{sec:model} gives the detailed description of our problem. The heuristic algorithm is presented in Section~\ref{sec:algo}.In Section~\ref{sec:exp}, we implement the algorithm with several instances of actual sizes and show experimental results. Finally, the paper is concluded ib Section~\ref{sec:conc}.

\section{Model}\label{sec:model}

The formal definition of our problem is given below. Let $G(V,E)$ be a directed graph. Each edge $e$ has a bandwidth capacity $c_e >0$ and a delay $d_e > 0$. We are given $k$ demands $\mathcal{D} = \{d_1, d_2, \ldots,d_k\}$, where $d_i=(s_i, t_i) \in V \times V$, along with three functions $\mathfrak{band}: \mathcal{D} \rightarrow R^+$, $\mathfrak{delay}: \mathcal{D} \rightarrow R^+$. $\mathfrak{hop}: \mathcal{D} \rightarrow R^+$. $\mathfrak{delay}(d_i)$ is the delay constraint for the demand $d_i$. The hop limit for each demand $d_i$ is $\mathfrak{hop}(d_i)$. Let $b_i = \mathfrak{band}(d_i)$ denote the bandwidth of demand $i$.

A simple $s_i$-$t_i$ path $p$ is feasible for the demand $d_i$ if it satisfies the delay and hop constraints:
\begin{equation}\label{ieq:delay}
\sum_{e \in p} d_e \leq \mathfrak{delay}(d_i)
\end{equation}

and
\begin{equation}\label{ieq:hop}
\sum_{e \in p} 1 \leq \mathfrak{hop}(d_i).
\end{equation}

A demand set $Q \subseteq \mathcal{D}$ is called {\em satisfied} if we can arrange a feasible path $p(d)$ for each demand $d$ in $Q$ and the total bandwidth usage on each edge $e$ does not exceed its capacity $c_e$. Namely, for any
$e\in G$ and $d \in Q$,
\begin{equation}\label{ieq:band}
\sum_{d :\, e \in p(d) } \mathfrak{band}(d) \leq c_e.
\end{equation}

The bandwidth of a demand set is the total bandwidth of all demands in the set. The goal of our problem is to find a feasible demand set $Q$ with a maximum bandwidth. Note that for each demand, the constraints~\eqref{ieq:delay} and~\eqref{ieq:hop} are ``local constraints'', which are irrelevant to other demands. However, the constraint~\eqref{ieq:band} is a ``global constraint'', where different demands might affect each other. Usually, a global constraint is much harder to handle than the local ones.

Recall that $b_i=\mathfrak{band}(d_i)$ is the bandwidth of demand $d_i$. In order to address an integer program more concisely, let $\mathcal P_i$ represent the set of all feasible paths between the source and destination of demand $d_i$. And the decision variable  $x_{ij} = 1$ indicates that the $j$-th path $p_{ij} \in \mathcal P_i$ is assigned for $d_i$, and $x_{ij} = 0$, otherwise. We can formulate the bandwidth-delay-hop constrained routing problem as below:

\begin{equation}\label{ilp:1}
\begin{aligned}
&  \text{max}
& & \sum_{i=1}^{k} \sum_{j=1}^{|\mathcal P_i|} b_i x_{ij}  &(IP)\\
& \text{s.t.} \\
& & & \sum_{p_{ij}: e \in p_{ij}} x_{ij} b_i \leq c_e   &   \forall  e \in E \\
& & & \sum_{j = 1}^{|\mathcal P_i|}x_{ij} \leq 1                 &  \forall i \in \{ 1,2,...,k \}  \\
& & & x_{ij} \in \{ 0,1 \}                              &  \forall i,j\\
\end{aligned}
\end{equation}

As $\mathcal P_i$ can be exponentially large, it is impossible to carry out all paths for each demand in a network of medium sizes. On the other hand, even if $\mathcal P_i$ is given, the integer program cannot be solved optimally either. The number of variables is too large for the existing general integer program solvers to get a solution in a few seconds.

To see the hardness of solving the above integer program, let us focus on a simplest variant where each $\mathcal P_i$ consists of exactly one path with a bandwidth of one, and each edge holds bandwidth capacity one as well. Then the problem is reduced to seeking a maximum number of edge-disjoint paths among the given path sets. Such a variant is called the path set packing problem~\cite{DBLP:conf/cocoon/XuZ18}, which is already NP-hard even if all given paths are of hop at most three. Moreover it is shown to be as hard as the well-known set packing problem in terms of approximation. Namely it cannot be approximated within a factor of $O(m^{1/2-\epsilon})$ for any $\epsilon>0$ unless $NP=ZPP$, where $m$ is the number of edges in the network.

The negative results in~\cite{DBLP:conf/cocoon/XuZ18} rule out any non-trivial theoretical guarantee on approximation algorithms. It motivates us to pay attention to heuristic approaches, which can run efficiently and effectively for large instances in practice.

\section{The Proposed Algorithm}\label{sec:algo}
In this section, a fast and efficient heuristic algorithm for the bandwidth-delay-hop constrained routing problem is proposed. We first describe the algorithmic framework, and then discuss in detail the individual components.

\subsection{Outline of The Algorithm}

Recall that the previously known routing algorithms deal with demands sequentially and combine path computing with path selection in one round. Such an idea does not work for our problem. Inspired by the integer program~(\ref{ilp:1}), we separate the path computing from the path selection. Let us first consider $\mathcal P_i$. It definitely not necessary to find all feasible paths for a demand, while a few paths may not suffice. Thus, we need to derive an efficient algorithm to generate a small but nice portion $P_i$ of $\mathcal P_i$  for each demand $d_i$. It is worthy noting that in path computing, the process for each demand is independent, which is allowed to run in parallel.

Fig.~\ref{algo:all} presents the approach, which runs iteratively. In each iteration, the algorithm consists of three phases: path computing, demand sorting, and path selection. In Phase 1, for each unsatisfied demand, it computes a number of candidate paths in parallel, satisfying the delay and hop constraints. In Phase 2, it sorts all unsatisfied demands with some rules. In Phase 3, the algorithm deals with the sorted demands one by one and tries to pick a feasible path from $P_i$ for each demand $d_i$. The network is updated accordingly as soon as a demand is assigned, i.e., the remaining capacity of each edge along the selected path decreases. A demand will be discarded in this iteration if none of its candidate paths is feasible. The algorithm (see Fig.~\ref{algo:all}) continues until none of the unsatisfied demands has a feasible path in a iteration.

\subsection{Specification of The Algorithm}\label{sec:salgo}

In this subsection, we will give the detailed description of each phase in the main algorithm, as well as the motivation behind. In Phase 1, we carry out the path computing problem with a bi-directed Breadth First Search (BFS). In Phase 2, four rules of demand sorting will be presented. These rules are designed based on different priorities. In Phase 3, we provide a mechanism for path selecting, which is motivated from the primal dual method.
\begin{figure}
	\caption{MAIN ALGORITHM}\label{algo:all}
	\begin{algorithmic}[1]
		
		\State \textbf{Input:} Network $G$ and demands $D=\{d_1 = (s_1,t_1),\cdots,d_k =(s_k,t_k)\}$
		
		\State $D_{satisfied}= \emptyset,\quad D_{unsatisfied}=D$
		
		\State $G_{residual}=G$
		
		\Repeat
		\State /* \emph{Phase 1: Path computing} */
		\ForAll {$i\in D_{unsatisfied}$}
		\State  $P_i \leftarrow path\_computing(G_{residual}, d_i)$
		\EndFor
		\State /* \emph{Phase 2: Demand sorting} */
		\State  Sort demands $D_{unsatisfied}$ according to a given rule.
		/* Detailed rules will be given in subsection~\ref{sec:rules}*/
		
		\State /* \emph{Phase 3: Path selecting} */
		\ForAll {$i\in D_{unsatisfied}$}
		\State  $P_i \leftarrow path\_selecting(G_{residual},d_i,P_i)$
		\If {$P_i \neq \emptyset$}
		\State  Pick the first path $p$ in the list of $P_i$. Let $D_{satisfied}\leftarrow D_{satisfied} \cup (d_i,p)$
		
		\State $D_{unsatisfied} \leftarrow D_{unsatisfied} \setminus d_i$
		\State Update graph $G_{residual}$, by setting the capacity of each edge $e \in p$ to be the current bandwidth $c_e$ minus $\mathfrak{band}(s_i,t_i)$, while keeping other parameters unchanged
		\EndIf
		
		\EndFor
		
		\Until {$|D_{satisfied}|$ does not increase}
		
		\State \textbf{Output:} $D_{satisfied}$		
	\end{algorithmic}
\end{figure}

\begin{figure}
	\caption{PATH\_COMPUTING$(G, d_i)$}
	\begin{algorithmic}[1]
		
		\State \textbf{Input:} one demand $d_i = (s_i,t_i)$, $\mathfrak{band} (d_i)$, $\mathfrak{delay}(d_i)$, $\mathfrak{hop}(d_i)$
		\State $P_i = \emptyset$
		\State Start a Breadth First Search (BFS) rooted at $s_i$. Let $d(v)$ denote the hop between $s_i$ and $v$ in the BFS tree. Stop the search if all vertices in $ S =  \{ v| d(v) \leq \lfloor \mathfrak{hop}(d_i) /2 \rfloor +1 \}$  have been visited.
		\State Start a reverse Breadth First Search (BFS) rooted at $t_i$. Let $d'(v)$ denote the hop between $t_i$ and $v$ in the BFS tree. Stop the search if all vertices in $ T =  \{ v| d'(v) \leq \lfloor \mathfrak{hop}(d_i) /2 \rfloor +1 \}$  have been visited.
		\For {each vertex $v$ visited by both Step 3 and Step 4. }
		\State Combine the $s_i$-$v$ path obtained from Step 3 and the $v$-$t_i$ path obtained from Step 4.
		\If {this new $s_i$-$t_i$ path is acyclic and satisfies the two constraints~(\ref{ieq:delay}) and (\ref{ieq:hop})}
		\State add it into $P_i$
		\EndIf
		\EndFor
		\State \textbf{Output:} $P_i$
	\end{algorithmic}\label{algo:pathc}
\end{figure}

\subsubsection{Paths Computing}

As mentioned above, we compute $P_i$ as a subset of all feasible paths for the demand $d_i$. In other words, we want to compute quite a lot of candidate paths for each demand independently. Our method does not consider the bandwidth constraints in this step, while the local constraints (on delay and hop) must hold.

Actually, there exists previous work in finding paths for a single demand with both delay and hop constraints. To the best of our knowledge, an algorithm called FindFP proposed in~\cite{Feng_Korkmaz_2015} is the fastest to compute feasible multi-constrained paths for a demand. However, as we pay attention to  large-scale networks with a large number of demands, the response time for such a subroutine must be very small. For instance, if the input is a network with $10,000$ nodes and $40,000$ edges, as well as $10,000$ demands, the designed algorithm needs to complete routing within 10 seconds. If we use FindFP to compute paths in this step, the running time is far more than the response time allowed according to Feng's experimental results\cite{Feng_Korkmaz_2015}. It tells that computing feasible multi-constrained paths directly is not doable. To make things work, we divide this phases into two steps: (a) compute paths that satisfy just one or even none of the constraints; (b) drop those paths violating one of the other constraints. Such a simple idea largely speeds up the process.

Fig.~\ref{algo:pathc} is used to compute paths for each demand.  We execute BFS from two opposite ways. Then feasible paths can be obtained by merging the half-way paths searched by BFS. Because of the hop-constraint, it is easy to see that any feasible path must pass one of the vertices that are visited by both BFS operations. In other words, if a vertex is not visited by either way of the BFS, no feasible path will be possible to reach this vertex. It follows that for any feasible path, there exists a path obtained in the algorithm whose length is not larger.

Different from paths obtained by the $k$-shortest paths algorithm, the paths returned by Fig.~\ref{algo:pathc} do not have many common edges with each other. The time complexity of Fig.~\ref{algo:pathc} is $O(|V|+|E|)$ if the hop is bounded by a constant, showing its efficiency in practice.

\subsubsection{Demand Sorting}\label{sec:rules}

After computing a candidate path set for each demand by Fig.~\ref{algo:pathc}, we can reconsider the integer program IP~(\ref{ilp:1}) using $P_i$ instead of $\mathcal P_i$. However, as discussed in the previous section, it is still impossible to tackle the program.

A naive idea is to limit the number of paths for each demand, namely, at most of a constant $k= 300$ paths are computing. Moreover, our algorithm decides to arrange the demands sequentially. In this case, the order of unsatisfied demands is crucial. We must be very careful in
sorting demands. Four rules are proposed below:
\begin{itemize}
	\item{\bf Rule 1.} Non-increasing lexicographical order of $(\mathfrak{band}(d_i),\frac{1}{\mathfrak{hop}(d_i)})$
	\item{\bf Rule 2.} Non-increasing lexicographical order of $(\frac{1}{\mathfrak{hop}(d_i)},\mathfrak{band}(d_i))$
	\item{\bf Rule 3.} Non-increasing order of $\frac{\mathfrak{band}(d_i)}{\mathfrak{hop}(d_i)}$
	\item{\bf Rule 4.} Non-decreasing order of $\mathfrak{hop}(d_i)*\mathfrak{band}(d_i)$.
\end{itemize}

In this work, we try all the four rules and return the maximum throughput. In the following, we will point out the intuitions for these rules.

\vskip 2mm\noindent
{\bf Rule 1} indicates that the demand which has a larger bandwidth should be considered earlier than the ones with smaller bandwidths, and the tie is broken in favor of the one whose hop is smaller. This rule is a natural greedy way in accordance to the objective maximizing the total bandwidth. Furthermore, if a demand has a larger hop, its path may contain more edges. With a larger possibility, this demand will occupy more network resource and may conflict with more other demands.

\vskip 2mm\noindent{\bf Rule 2} is similar to the first rule, while we give the priority to the hop-constraint other than the demand bandwidth.

\vskip 2mm\noindent{\bf Rule 3} is very similar to the rule mentioned in~\cite{Jozsa_2001}, where the demands are sorted by the decreasing order of its bandwidth divided by the minimum hop over the paths between its two endpoints. We do not apply their rule because it will take a lot of time to compute the minimum hop path for each demand.  Instead, we use the hop limit to replace the minimum hop.

\vskip 2mm\noindent
{\bf Rule 4} is motivated by mimicking the primal dual method to solve the integer program~(\ref{ilp:1}). If we relax the integer program~(\ref{ilp:1}) by replacing the constraints $x_{ij} \in \{0,1 \}$ with $x_{ij} \geq 0$, and taking its dual, we obtain

\begin{equation}
\begin{aligned}\label{lp:dual}
min &&\sum_{i=1}^{k} u_i + \sum_{e\in E} c_e v_e &&\\
s.t. && u_i + \sum_{e \in P_{ij}} b_i v_e \geq b_i & &\forall i,j \\
&& u_i \geq 0 && \forall i \in \{ 1,2,...,k \} \\
&&v_e \geq 0 && \forall e\in E
\end{aligned}
\end{equation}

Begining with a feasible dual solution: $u_i = b_i -1, \forall i \in \{1,...,k\}$ and $v_e = 1, \forall e \in E$, we decrease all $v_e$'s uniformly until $u_i + \sum_{e\in P_{ij}}b_i v_e = b_i$ for some $i, j$. The  path $p_{ij}$ is then chosen, which is corresponding to the demand $d_i$ with the least $b_i\cdot |p_{ij}|$, where $|p_{ij}|$ stands for the number of edges in the path $p_{ij}$. Noting that $\mathfrak{hop}_i*\mathfrak{band}_i$  is an upper bound of $b_i\cdot |p_{ij}|$,  it comes up with the fourth rule.

\vskip 2mm
\subsubsection{Path Selecting}

After sorting all unsatisfied demands, in this step, we will choose one path for each demand $d_i$ from its path set $P_i$. Recall that we ignore the bandwidth constraints when computing $P_i$. A path $p \in P_i$ violates the bandwidth constraint if one of the edges in $p$ exceeds its bandwidth capacity after assigning $p$. We first eliminate from $P_i$ those paths violating the bandwidth constraints. Then a weight is assigned to each of the remaining paths, based on which we can define the priority over the paths. The path of the minimum weight is taken. See Fig.~\ref{algo:pathselecting}.

Now we give an intuition of the weight assignment in Fig.~\ref{algo:pathselecting}. Recall the dual program~(\ref{lp:dual}). 
Replacing $c_ev_e$ with $v'_e$, we have:

\begin{equation}
\begin{aligned}\label{lp:pr}
min &&\sum_{i=1}^{k} u_i + \sum_{e\in E} v'_e &&\\
s.t. && u_i + \sum_{e \in P_{ij}} \frac{b_i}{c_e} v'_e \geq b_i & &\forall i,j \\
&& u_i \geq 0 && \forall i \in \{ 1,2,...,k \} \\
&&v'_e \geq 0 && \forall e\in E
\end{aligned}
\end{equation}

\begin{figure}
	\caption{PATH\_SELECTING$(G, d_i, P_i)$}
	\begin{algorithmic}[1]
		\State \textbf{Input:} a demand $d_i = (s_i,t_i)$ and $P_i$ that was computing in Fig.~\ref{algo:pathc}
		\ForAll {$p \in P_i$}
		\If{ p does not satisfy bandwidth constraints on $G_{residual}$ }
		\State $P_i \leftarrow P_i \setminus p$
		\Else
		\State $w(p):= \sum_{e \in p} \frac{1}{\text{residual bandwidth of e}}$
		\EndIf
		\EndFor
		\State Sort paths in $P_i$ in the non-decreasing order of of the weight $w(p)$.
		\State \textbf{Output:} $P_i$
	\end{algorithmic}\label{algo:pathselecting}
\end{figure}

Again we begin with a feasible dual solution $u_i  = b_i -1, \forall i \in \{1,...,k\}$ and $v'_e = \max_{e\in E} c_e, \forall e \in E$. With the similar operations, we find that the path choosing for the $i$-th demand is the one which has the least $\sum_{e\in P_{ij}} \frac{1}{c_e}$. Note that in each step, we rewrite the dual program, and $c_e$ is not the capacity of edge $e$ but the residual capacity. Thus, the weight $w(p)$ of path $p$ is defined as following: $w(p):= \sum_{e \in p} \frac{1}{\text{residual bandwidth of e}}$.
\vskip 2mm
This weight function shows that the more residual bandwidth a path has, the higher possibility this path is chosen. In other words, we choose the path that leaves as much space for other unsatisfied demands as possible.

\section{Experimental Results}\label{sec:exp}

We dedicate this section to experimental results for our algorithm. All experiments were conducted on a SUSE Linux 11 system with an E5-2690 v3 PC, which has 12 CPU cores and a total of $192G$ memory. During experiments, we use OpenMP to make parallel acceleration. The total number of threads is $48$.

To evaluate the performance of our algorithm, we use instances generated by Huawei technologies co. The criterion for generating instances is to simulate real networks with actual sizes. Basically, we make sure that at least 80\% demands can be satisfied in an optimal solution for each instance. The procedure is stated below.
\begin{enumerate}
	\item First, input the number $n$ of vertices, the number $m$ of edges, and the number $k$ of demands.
	
	\item Then, all vertices are randomly placed with a uniform distribution into a square area with a side length 100.
	
	\item Randomly choose two vertices. If their Euclidean distance is less than 80, connect them with an edge, whose delay is a random value in between 50 and 100.
	
	\item Repeat Step 3 until the number of edges reaches $m$.
	
	\item Randomly choose two vertices as a demand with $\mathfrak{band}$  between 1,000 and 5,000. Find an arbitrary path connecting them, which is denoted as a pre-selected path. Let the total delay and the length of the pre-selected path be the demand's $\mathfrak{delay}$ and $\mathfrak{hop}$, respectively.
	
	\item Repeat Step 5 until the number of demands reaches $k$.
	
	\item Randomly choose $k*80\%$ demands generated in Step 6. Now we define a bandwidth capacity $c_e$ for each edge $e$. Summing up $\mathfrak{band}$ of the demands whose selected path passes the edge $e$, and multiplying it by $1.25$, we get $c_e$.
\end{enumerate}

Our algorithms are tested in two different scenarios: medium size networks and large-scaled networks. We generate \textbf{Instances A} for a medium size network, which is a set of $100$ instances that have $10,000$ demands and a network with $500$ nodes and $2,000$ edges. \textbf{Instances B} is generated for a large-scaled network, which have $10,000$ demands, a network with $10,000$ nodes and $40,000$ edges. The number of instances in \textbf{Instances B} is also $100$.

Corresponding to the three phases in our algorithm, we design three experiments to illustrate the performance of the algorithm. The first experiment compares our algorithm with those approaches in which Fig.~\ref{algo:pathc} is replaced by some other known algorithms. In the second experiment, we implement our algorithm under the four demands sorting rules, and compare with each other. The third experiment is designed to show the path selecting rule is powerful.
The three experiments will present an evidence that our algorithm is superior in each phase.

In addition, we also compare our algorithm with three existing routing algorithms to convince that our algorithm performs really much better. More details of each experiment will be provided in the following.

\subsection{Experiment 1}
The most important part of our algorithm is the candidate path computing for each demand. In this section, we compare our method of computing candidate paths with the well-known $k$-shortest path algorithm. To highlight the difference, we call it the \emph{kSPA} algorithm if we replace Fig.~\ref{algo:pathc} with a $k$-shortest path algorithm. In this experiment, we adopt the current fastest $k$-shortest path algorithm by Eppstein~\cite{Eppstein_1998}.
The $k$-shortest path algorithm computes $k$ shortest paths between a source node and a destination node with respect to a given specific weight on every edge. The throughput of the \emph{kSPA} algorithm depends on the choice of $k$. The larger $k$ is, the higher throughput one can expect. However, if $k$ is too large, the running time is unacceptable. In this experiment, let $k = 128$. We also only keep $128$ paths in Fig.~\ref{algo:pathc}.

For both our algorithm and the \emph{kSPA} algorithm, we adopt Rule~1 in Phase 2 and Fig.~\ref{algo:pathselecting} in Phase 3.

Table~\ref{tb:t-exp1} presents the running time for our algorithm and the \emph{kSPA} algorithm, respectively. It shows that our algorithm is more than $20$ times faster.

Our algorithm is not only faster than the \emph{kSPA} Algorithm, but the average throughput given by our algorithm is also around 5\% higher than the one given by the \emph{kSPA} Algorithm. See details in Table~\ref{tb:thr-exp1}. For both instances $A$ and $B$, our algorithm achieves a very high throughput, which are 97.71\% and 92.55\% in average, respectively.

\begin{table}
	\centering
	\caption{Comparison of Running Times}\label{tb:t-exp1}
	\begin{tabular}{cl|ccc}
		\hline
		\multicolumn{1}{l}{\textbf{}}                                                                                         &      & \begin{tabular}[c]{@{}c@{}}Our\\ Algorithm\end{tabular} & \begin{tabular}[c]{@{}c@{}} the \emph{kSPA}\\ Algorithm\\ (delay set \\ as weight)\end{tabular} & \begin{tabular}[c]{@{}c@{}}the \emph{kSPA}\\ Algorithm\\ (hop set\\ as weight)\end{tabular} \\ \hline
		\multicolumn{1}{c|}{\multirow{4}{*}{\begin{tabular}[c]{@{}c@{}}Running\\ time (sec)\\ on Instances A\end{tabular}}} & avg. & 0.57                                                    & 11.03                                                                          & 10.93                                                                        \\
		\multicolumn{1}{c|}{}                                                                                                 & s.d. & 0.12                                                    & 1.85                                                                           & 2.07                                                                         \\
		\multicolumn{1}{c|}{}                                                                                                 & max. & 0.81                                                    & 16.68                                                                          & 15.24                                                                        \\
		\multicolumn{1}{c|}{}                                                                                                 & min. & 0.39                                                    & 8.76                                                                           & 7.42                                                                         \\ \hline
		\multicolumn{1}{c|}{\multirow{4}{*}{\begin{tabular}[c]{@{}c@{}}Running\\ time(sec)\\ on Instances B\end{tabular}}}  & avg. & 5.82                                                    & 162.20                                                                         & 155.72                                                                       \\
		\multicolumn{1}{c|}{}                                                                                                 & s.d. & 1.75                                                    & 46.18                                                                          & 45.28                                                                        \\
		\multicolumn{1}{c|}{}                                                                                                 & max. & 8.77                                                    & 225.69                                                                         & 221.41                                                                       \\
		\multicolumn{1}{c|}{}                                                                                                 & min. & 3.18                                                    & 85.90                                                                          & 82.05                                                                        \\ \hline
	\end{tabular}
\end{table}

\begin{table}
	\centering
	\caption{Comparison of Throughputs}\label{tb:thr-exp1}
	\begin{tabular}{cl|ccc}
		\hline
		\multicolumn{1}{l}{\textbf{}}                                                                                                        &      & \begin{tabular}[c]{@{}c@{}}Our\\ Algorithm\end{tabular} & \begin{tabular}[c]{@{}c@{}}the \emph{kSPA}\\ Algorithm\\ (delay set\\ as weight)\end{tabular} & \begin{tabular}[c]{@{}c@{}}the \emph{kSPA}\\ Algorithm\\ (hop set\\ as weight)\end{tabular} \\ \hline
		\multicolumn{1}{c|}{\multirow{4}{*}{\begin{tabular}[c]{@{}c@{}}Throughput:\\ \% of total\\ bandwidth\\ on Instances A\end{tabular}}} & avg. & 95.58                                                   & 90.44                                                                          & 90.92                                                                        \\
		\multicolumn{1}{c|}{}                                                                                                                & s.d. & 3.10                                                    & 7.89                                                                           & 6.86                                                                         \\
		\multicolumn{1}{c|}{}                                                                                                                & max. & 100                                                     & 99.93                                                                          & 99.94                                                                        \\
		\multicolumn{1}{c|}{}                                                                                                                & min. & 78.95                                                   & 72.06                                                                          & 70.11                                                                        \\ \hline
		\multicolumn{1}{c|}{\multirow{4}{*}{\begin{tabular}[c]{@{}c@{}}Throughput:\\ \% of total\\ bandwidth\\ on Instances B\end{tabular}}} & avg. & 92.55                                                   & 85.67                                                                          & 87.22                                                                        \\
		\multicolumn{1}{c|}{}                                                                                                                & s.d. & 5.01                                                    & 9.80                                                                           & 8.49                                                                         \\
		\multicolumn{1}{c|}{}                                                                                                                & max. & 99.74                                                   & 97.6                                                                           & 98.27                                                                        \\
		\multicolumn{1}{c|}{}                                                                                                                & min. & 78.71                                                   & 63.4                                                                           & 69.09                                                                        \\ \hline
	\end{tabular}
\end{table}

\subsection{Experiment 2}

In this subsection, we execute our algorithm under the four rules for demand sorting. In addition, we consider the case where we do not sort demands. Table~\ref{tb:thrs} shows the average (avg.), standard deviation (s.d.), maximum (max.) and minimum (min.) of the throughput in two types of instances, respectively. We can see that the four rules are not of significant differences. All of them achieve a high throughput and their standard deviations are all smaller. This implies that it is necessary to sort demands with the rules.


\begin{table*}[htbp]
	\centering
	\caption{Comparison of Sorting Rules}\label{tb:thrs}
	\begin{tabular}{cl|ccccc}
		\hline
		\textbf{}                                                                                                                            & \multicolumn{1}{c|}{} & \emph{Rule 1} & \emph{Rule 2} & \emph{Rule 3} & \emph{Rule 4} & \begin{tabular}[c]{@{}c@{}}No sorting\end{tabular} \\ \hline
		\multicolumn{1}{c|}{\multirow{4}{*}{\begin{tabular}[c]{@{}c@{}}Throughput:\\ \% of total\\ bandwidth\\ on Instances A\end{tabular}}} & avg.                  & 97.71  & 97.41  & 97.64  & 97.72  & 95.21                                                       \\
		\multicolumn{1}{c|}{}                                                                                                                & s.d.                  & 3.29   & 3.62   & 3.38   & 3.28   & 8.29                                                        \\
		\multicolumn{1}{c|}{}                                                                                                                & max.                  & 100    & 100    & 100    & 100    & 100                                                         \\
		\multicolumn{1}{c|}{}                                                                                                                & min.                  & 88.85  & 88.35  & 88.45  & 88.77  & 72.85                                                       \\ \hline
		\multicolumn{1}{c|}{\multirow{4}{*}{\begin{tabular}[c]{@{}c@{}}Throughput:\\ \% of total\\ bandwidth\\ on Instances B\end{tabular}}} & avg.                  & 92.55  & 90.17  & 92.02  & 92.54  & 90.27                                                       \\
		\multicolumn{1}{c|}{}                                                                                                                & s.d.                  & 6.31   & 9.78   & 7.00   & 6.30   & 9.77                                                        \\
		\multicolumn{1}{c|}{}                                                                                                                & max.                  & 99.74  & 99.79  & 99.77  & 99.74  & 99.79                                                       \\
		\multicolumn{1}{c|}{}                                                                                                                & min.                  & 80.71  & 64.72  & 78.37  & 70.71  & 65.95                                                       \\ \hline
	\end{tabular}
\end{table*}

\subsection{Experiment 3}

Different ways of selecting paths certainly result in a different throughput. In this experiment, for each demand, we try three other methods instead: (1) Choose the path with the smallest hop. (2) Choose the path with the smallest delay. (3) Randomly choose a feasible path in this phase. Like Experiment 1, we use Rule 1 in phase 2. Table~\ref{tb:p-exp3} shows that our path selecting algorithm~\ref{algo:pathselecting} performs better than the others. It is important to take account into the residual bandwidth of the edges.

\begin{table*}[htbp]
	\centering
	\caption{Comparison of Path Selecting}\label{tb:p-exp3}
	\begin{tabular}{cl|cccc}
		\hline
		\textbf{}                                                                                                                            & \multicolumn{1}{c|}{} & \begin{tabular}[c]{@{}c@{}}Our\\ Algorithm\end{tabular} & \begin{tabular}[c]{@{}c@{}}The least\\ hop path\end{tabular} & \begin{tabular}[c]{@{}c@{}}The least\\ delay path\end{tabular} & \begin{tabular}[c]{@{}c@{}}Random\\ feasible path\end{tabular} \\ \hline
		\multicolumn{1}{c|}{\multirow{4}{*}{\begin{tabular}[c]{@{}c@{}}Throughput:\\ \% of total\\ bandwidth\\ on Instances A\end{tabular}}} & avg.                  & 97.71                                                   & 95.20                                                        & 91.56                                                          & 94.50                                                          \\
		\multicolumn{1}{c|}{}                                                                                                                & s.d.                  & 3.29                                                    & 5.97                                                         & 8.75                                                           & 6.38                                                           \\
		\multicolumn{1}{c|}{}                                                                                                                & max.                  & 100                                                     & 100                                                          & 99.48                                                          & 99.88                                                          \\
		\multicolumn{1}{c|}{}                                                                                                                & min.                  & 88.85                                                   & 79.7                                                         & 73.99                                                          & 79.5                                                           \\ \hline
		\multicolumn{1}{c|}{\multirow{4}{*}{\begin{tabular}[c]{@{}c@{}}Throughput:\\ \% of total\\ bandwidth\\ on Instances B\end{tabular}}} & avg.                  & 92.55                                                   & 89.25                                                        & 87.49                                                          & 88.34                                                          \\
		\multicolumn{1}{c|}{}                                                                                                                & s.d.                  & 6.31                                                    & 7.40                                                         & 7.79                                                           & 7.47                                                           \\
		\multicolumn{1}{c|}{}                                                                                                                & max.                  & 99.74                                                   & 99.31                                                        & 98.00                                                          & 98.84                                                          \\
		\multicolumn{1}{c|}{}                                                                                                                & min.                  & 80.71                                                   & 76.91                                                        & 75.35                                                          & 76.91                                                          \\ \hline
	\end{tabular}
\end{table*}

\subsection{Experiment 4}

In the last experiment, we compare our algorithm (using Rule 1) with three other algorithms: the minimum delay  algorithm~\cite{wang_crowcroft_1996}(MDA), the widest shortest path algorithm~\cite{qoS_routingmechanisms_1999}(WSP), and the shortest widest path algorithm~\cite{wang_crowcroft_1996}(SWP). Since the real network is of a large scale, it costs a lot to find the shortest paths with respect to delay for all ingress-egress node pairs. So in this experiment, we do not apply the algorithms proposed in~\cite{Jozsa_2001} and~\cite{tomovic_radusinovic_2016}.
To ensure the paths computed by MDA, WSP and SWP satisfy the hop and delay constraints, we replace the shortest path algorithm with the extended Dijkstra shortest path algorithm\cite{Chen_1999}.

\begin{table}
	\centering
	\caption{Comparison of Four Algorithms}\label{tb:p-exp4}
	\begin{tabular}{cl|cccc}
		\hline
		\textbf{}                                                                                                                            & \multicolumn{1}{c|}{} & \begin{tabular}[c]{@{}c@{}}Our\\ Algorithm\end{tabular} & \emph{MDA}   & \emph{WSP}   &  \emph{SWP}   \\ \hline
		\multicolumn{1}{c|}{\multirow{4}{*}{\begin{tabular}[c]{@{}c@{}}Throughput:\\ \% of total\\ bandwidth\\ on Instances A\end{tabular}}} & avg.                  & 97.71                                                   & 73.99 & 77.20 & 69.17 \\
		\multicolumn{1}{c|}{}                                                                                                                & s.d.                  & 3.29                                                    & 18.20 & 17.18 & 9.38  \\
		\multicolumn{1}{c|}{}                                                                                                                & max.                  & 100                                                     & 96.88 & 98.16 & 82.60 \\
		\multicolumn{1}{c|}{}                                                                                                                & min.                  & 88.85                                                   & 43.32 & 46.18 & 51.23 \\ \hline
		\multicolumn{1}{c|}{\multirow{4}{*}{\begin{tabular}[c]{@{}c@{}}Throughput:\\ \% of total\\ bandwidth\\ on Instances B\end{tabular}}} & avg.                  & 92.55                                                   & 47.85 & 52.06 & 50.89 \\
		\multicolumn{1}{c|}{}                                                                                                                & s.d.                  & 6.31                                                    & 18.40 & 18.19 & 12.06 \\
		\multicolumn{1}{c|}{}                                                                                                                & max.                  & 99.74                                                   & 83.97 & 85.68 & 69.64 \\
		\multicolumn{1}{c|}{}                                                                                                                & min.                  & 80.71                                                   & 21.82 & 25.33 & 32.32 \\ \hline
	\end{tabular}
\end{table}

From Table~\ref{tb:p-exp4} it is obvious that our algorithm runs much better than the other algorithms. We think it is owing to the advantage of computing several paths from each demand, while all the three other algorithms deal with demands one by one and just compute one feasible path for each demand. The average of SWP's throughput is slightly less than the throughput for MDA and WSP, but SWP performs much more stably than them. This phenomenon indicates that the residual capacity of the edges on a path should be considered more preferentially than the hop or delay of the path if we take care of the robustness of the algorithm.

Moreover, the three algorithms cannot be executed in parallel, which leads to the situation that the running time increases rapidly when the network becomes larger. Their average running time for Instances B is nearly one and half an hour, while our algorithm only takes 5.82 seconds on average.

\section{Concluding Remarks}\label{sec:conc}
We have studied the routing algorithm with bandwidth-hop-delay constraints, which can be applied in an SDN architecture. Our proposed algorithm is fast, which is within 10 seconds, for a large-scaled network with around 10,000 nodes and 40,000 edges. And it also outputs a throughput around 97\% and 92\% for medium scale network and large scale network, respectively.

Many problems are open in this line of research, since different practical scenarios may suggest different constraints. In addition to delay and hop constraints, one may need to find a path which has to visit some given vertices or has to avoid some given vertices. Carrying out backup routing paths for each demand is an interesting topic as well.

\section*{Acknowledgments}

The authors would like to thank the Huawei Technologies Company for providing such a challenging problem and suggesting a way to generate experimental data. The fourth author is supported in part by the National Natural Science Foundation of China under Grant No. 11671355.

\bibliographystyle{plain}
\bibliography{sdn.bib}

\end{document}